\documentclass[11pt]{article}%
\usepackage{makeidx, siunitx, ragged2e}
\usepackage{booktabs,lscape,color, float}
\usepackage{subcaption} 
\usepackage{amsfonts,bm}
\usepackage{amsmath}
\usepackage{amssymb}
\usepackage{graphicx}
\usepackage{rotating}
\usepackage{multirow}%
\usepackage{geometry,setspace}
\usepackage{natbib}
\usepackage{url}
\date{}
\begin{document}
	
	\title{A New Look to Three-Factor Fama-French Regression Model using Sample Innovations}
	\author{
		Javad Shaabani\thanks{Department of Statistics, Yazd University, Yazd, Iran,  javadshaabani@stu.yazd.ac.ir.}
			\and Ali Akbar Jafari\thanks{Department of Statistics, Yazd University, Yazd, Iran, aajafari@yazd.ac.ir.} 
	}
	\maketitle

	\begin{abstract}
The Fama-French model is widely used in assessing the portfolio's performance compared to market returns. In Fama-French models, all factors are time-series data. The cross-sectional data are slightly different from the time series data. A distinct problem with time-series regressions is that R-squared in time series regressions are usually very high, especially compared with typical R-squared for cross-sectional data. The high value of R-squared may cause misinterpretation that the regression model fits the observed data well, and the variance in the dependent variable is explained well by the independent variables. Thus, to do regression analysis, and overcome with the serial dependence and volatility clustering, we use standard econometrics time series models to derive sample innovations. In this study, we revisit and validate the Fama-French models in two different ways: using the factors and asset returns in the Fama-French model and considering the sample innovations in the Fama-French model instead of studying the factors. Comparing the two methods considered in this study, we suggest the Fama-French model should be consider with heavy tail distributions as the  tail behavior is relevant in Fama-French models, including financial data, and the QQ plot does not validate that the choice of normal distribution as the theoretical distribution for the noise in the model.
		
	\end{abstract}
	
\noindent\textbf{Keywords:} Fama-French model; asset return modeling; sample innovations; econometrics.  
	
	\newpage
	\section{Introduction}\label{sec:intro}
	
	\cite{Sharpe:1964} (Nobel laureate in economics) and \cite{Lintner:1965} introduced the capital asset pricing model (CAPM) to  measure the investment risk and the return of an individual stock. The CAPM gives sturdy and intuitively pleasing forecasts about how to asses risk and the relation between risk and expected return. 
	Although the expected return of the market and market risk are the two factors taking into account in the CAPM model, the CAPM model employs only one variable, market return, to explain the stock returns. The standard formula of the CAPM is 
	\begin{equation}
	\label{CAPM}
	r_{t}-r_{f,t}=\alpha+\beta\left( r_{m,t}-r_{f,t}\right)+\epsilon_t 
	\end{equation} 
	where $r$ is return on an asset; $r_f$ is risk-free rate; $r_m$ is the return of the market; $\beta$ is the sensitivity of the  asset returns to the excess market returns $\left( r_m-r_f\right)$; $\alpha$ is the unexplained expected return by asset; and  $\epsilon $ is the market noise (error term associated with the excess return of the asset). 
	
	The CAPM was widely used in assessing the portfolio's performance compared to market returns and measuring risk, beta, and alpha among practitioners. Many academic studies have determined uncertainty on reliability of the CAPM.  \cite{FAMA:1992} by studying the share returns on the American Stock Exchange, the New York Stock Exchange, and Nasdaq \footnote{See \url{https://www.nasdaq.com/}.} observed that differences in $\beta$ over a lengthy period did not describe the performance of different stocks. Also, the linear relation between stock returns and $\beta$ breaks down over shorter periods. These findings imply the weaknesses of the CAPM model in applications. 
	
	\cite{Fama:1993} extended the CAPM model to create a better tool to evaluate portfolio performance and to predict the stock returns  by adding two more factors to the CAPM. 
	 The factors in Fama-French model  are (1) market risk, (2) the outperformance of small-cap companies over large-cap companies, and (3) the excess returns of high book-to-price\footnote{The book-to-bill ratio is the ratio of orders received to the amount billed for a specific period, usually one month or one quarter. See \url{https://en.wikipedia.org/wiki/Book-to-bill_ratio}.} ratio companies versus small low book-to-price ratio companies. The reason behind the Fama-French three-factor model is that high value and small-cap companies tend to beat the overall market usually. The Fama-French three-factor's representation model is
	 
	 \begin{equation}
	 \label{FF3}
	 r_{t}-r_{f,t}= \alpha + \beta_1\left( r_{m,t}-r_{f,t}\right)+ \beta_2 SMB_t+
	 \beta_3 HML_t+
	 \epsilon_t 
	 \end{equation} 
	 where $SMB$ stands for ``Small market capitalization Minus Big market capitalization'' and $HML$ for ``High book-to-market ratio Minus Low book-to-market ratio;'' they measure the historic excess returns of small caps over big caps and of value stocks over growth stocks.

	 In Fama-French models, all factors are time-series data. The cross-sectional data are slightly different from the time series data. 	 A distinct problem with time-series regressions is that R-squared in time series regressions are usually very high, especially compared with typical R-squared for cross-sectional data. However, it does not imply that we learn more about factors impacting response variable from time-series data.  One reason for artificially high values of R-squares and adjusted R-squares for time series regressions is because the response variable and dependent variables have trends over time. A high R-squared can occasionally indicate that there is a problem with your model. It may cause misinterpretation that the regression model fits the observed data well, and the variance in the dependent variable is explained well by the independent variables.

	 Another distinct aspect of time series data that identifies it from cross-sectional data is that a time series data set arrives with a temporal ordering. In other words, the past can impact   the future. 
	 Generally, time-series data drops random sampling, Endogeneity, and homoscedasticity assumptions of the Gauss-Markov Theorem.  To do regression analysis for time series data, we need to make two assumptions (1) the expected value of $\epsilon_t$ is not correlated with the explanatory variables in any periods and (2) $\epsilon_t$ is also  uncorrelated with its past and future values (no autocorrelation). Thus we should text these assumptions before performing regression analysis.

	The hypotheses of independent and identically distributed for $r_t$, $r_f$, $SMB$, and $HML$ are rejected. The Ljung-Box Q-test \citep[]{ljung:1978} indicates the presence of autocorrelation and heteroskedasticity in data set. The Engle-Granger co-integration \citep[]{fuller:2009} test rejects the null hypothesis of no co-integration among the time series. 
	The explanatory variables drop the random sampling, endogeneity, and homoscedasticity assumptions in the Gauss-Markov Theorem. 
	
    Therefore, to do regression analysis, and overcome with the serial dependence and volatility clustering, we use standard econometrics time series models to derive sample innovations. Instead of studying the factors and asset returns, we consider their sample innovations derived from the time series models. Consequently, we have iid standardized residuals for each factor as explanatory variables in the Fama-French models.

	 Therefore, in this study, we revisit the Fama-French models in two different ways:
	 \begin{itemize}
	 	\item Using the factors and asset returns in the Fama-French model and validating the model by ruing time series regression;
	 	\item Filtering linear and  nonlinear temporal dependencies in factors by applying time series model. Instead of studying the factors and asset returns in the model, considering the sample innovations for validating the Fama-French model; and 
	 \end{itemize}

	  We use Microsoft's stock data to test  the robustness of the Fama-French (FF) models in
	 explaining the variation in stock returns. The reason behind of choosing Microsoft  stock is that Microsoft is the largest publicly traded US company and the top constituent of the S\&P 500 index.

	 The results from the QQ plot show the frequency of extreme events is higher than that implied by the normal distribution. These result indicate that the distribution of the OLS regression estimator does not acquire the tail behavior. Tail behavior is relevant for regressions, including financial data. A potential reason for the significant variation in the estimated regression coefficients across time is the heavy-tailed nature of the distribution of the innovations. It is an approved empirical fact that many financial variables are modeled better by distributions with tails heavier than the normal distribution. 	 The result of the two methods considered in this study, suggest the Fama-French model should be consider with heavy tail distributions. 

	We note that there is a comprehensive research that has attempted to model the tail behavior of asset returns and to deal with non-normality of asset return.   Modeling the tail behaviors of asset returns is essential for risk managers. The the method of subordination \citep[see][]{Sato:2002} widely proposed in literature
	to introduce additional parameters to the return model to reflect the heavy tail phenomena present in most asset returns and to generalize the classical asset pricing model \citep[see][]{Clark:1973,Barndorff:1977,Carr:2004,Klingler:2013,shirvani2020multiple}. To  incorporate the views of investors 	into asset return models and to deal with non-normality of asset return, \cite{shirvani2020option} a new process for asset returns in the form of a mixed geometric Brownian motion and subordinated Levy process. In this study, to model the asset return, we propose the  Fama-French model with heavy tail distributions as the theoretical distribution for the noise in the model.

	There are three sections that follow in this paper. In the next section, Section 2, we describe the data source and data validation. Our methodology for modeling time series data is described in Section 3. Section 4 compares the two methods used in this study. The results of each method are reported in two subsections. Section 5 concludes the paper.

	\section{Data source and data validation}

	To examine the robustness of the Fama-French (FF) model in
	explaining the variation in stock returns, we use the  Microsoft \footnote{\url{	https://www.microsoft.com/en-us/investor}.} stock price data  as risky asset. The 10-year Treasury yield is used as returns of the riskless asset.\footnote{\url{https://ycharts.com/indicators/10_year_treasury_rate}.} 
	The  historical data for factors  in Fama-french model (SMB, HML, RMW, and CMA) are collected from the Kenneth French-Data Library.\footnote{\url{https://mba.tuck.dartmouth.edu/pages/faculty/ken.french/data_library.html}.}
	The data set are monthly in the period from om March 1986 to February 2020. 	
	The factors in the Fama-French model with the abbreviated
	names are  follows.\footnote{See \url{https://corporatefinanceinstitute.com/resources/knowledge/finance/}\\  \url{fama-french-three-factor-model/}.} 

	\begin{description}
	\item[\texttt{Market risk premium}.] Market risk premium (MRP) is the difference between the expected return of the market and the risk-free rate. It provides an investor with an excess return as compensation for the additional volatility of returns over and above the risk-free rate.
	
	\item[\texttt{Small Minus Big}.] Small Minus Big (SMB) is a size effect based on the market capitalization of a company. SMB measures the historic excess of small-cap companies over big-cap companies.
	
	\item[\texttt{High Minus Low}.] High Minus Low (HML) is a value premium. It represents the spread in returns between companies with a high book-to-market value ratio (value companies) and companies with a low book-to-market value ratio. 
	
\end{description}

	\section{Methods}
	 The most important of the functional forms that we use in time series regressions is the natural logarithm.  As  the time series data are not statistically significant, we account this change to remove a time trend. 
	In our analysis, the returns of the risky asset are obtained by the 
	logarithmic return  as follows
	\begin{equation}
	\label{log-return}
	r(t)\,\,=\,\, \ln \frac{S(t)}{S(t-1)}, 
	\end{equation}
	where $S(t)$ is the price of Microsoft stock at day $t$. 
	
	The \textit{Autoregressive Moving Average} (ARMA) \citep[]{fuller:2009}
	 and the \textit{Autoregressive Conditional Heteroscedasticity} (GARCH) model \citep[]{Hamilton:1994} are the standard tools for modeling the mean and volatility of times series data set. 
	 Here, to filter out the trend and dependence in data set, we use the  ARMA$(1,1)$--GARCH$(1,1)$ with normal distribution assumption for innovations as follows 
	\begin{equation}
	\label{arma-garch}
	\left\{
	\begin{array}{lcl}
	r_t & =& \mu+\varphi (r_{t-1}-\mu)+\theta a_{t-1}+a_t, \\
	a_t& =&\varepsilon_t \sigma_t,  \,\,\, \varepsilon_t  \sim  \mathrm{iid,} \\
	\sigma_t^2 &= & \gamma+ \alpha a_{t-1}^2+\beta\sigma_{t-1}^2,
	\end{array}
	\right.
	\end{equation}
	$\mu_t$ and $\sigma_t$ are the conditional mean and volatility of returns for stock and bonds,   
	$\varepsilon_t$ is standardized \textit{iid}--normal random variable, $a_t$ is referred to the market shocks, 
	$ \alpha\geq 0, \beta\geq 0, \gamma\geq 0, \delta,\varphi,\theta $ are the model parameters. We fit ARMA$(1,1)$--GARCH$(1,1)$ with normal distribution assumption for innovations on each factor. The model parameters are estimated  with the use of the R-Package \textit{``rugarch''} \citep[]{ghalanos:2019}. Instead of using the factors and asset returns in the Fama-French model, we consider the sample innovations obtained from each time series model.

	\section{Empirical Results}
	\subsection{Fama-French three-factor model summary statistics}
	 The following time-series regression (Fama-French 3 factor model) is used in this section: 
	 \begin{equation*}
		r_{t}-r_{f,t}= \alpha + \beta_1\left( r_{m,t}-r_{f,t}\right)+ \beta_2 SMB_t+
	\beta_3 HML_t+
	\epsilon_t 
	\end{equation*} 
	
	As to conduct OLS stationary test is required, we used the Augmented Dickey-Fuller.  The p-values$\left(<0.01 \right)$ are about zero indicating stationary time series data set. As the data was transformed into the log-return, we expected to have a stationary time series.  \cite{brooks:2019} explained that non-stationary data is led to spurious regressions. Table 1 shows the average monthly rate of return for Microsoft stock and the standard deviation for factors. As we observed from Table 1, the average monthly rate of return for market portfolio is higher than the Microsoft stock. It indicates the market outperforms the big market capitalization stocks. The reason behind it is that the companies with small market capitalization perform better than the significant market capitalization portfolios.

\begin{table}[h!]
	\centering
	\begin{tabular}{@{}lcccccc@{}}
		\toprule
		& EXR   & MRP    & SMB    & HML    & RMW    & CMA    \\ \midrule
		Mean     & 0.0055 & 0.0065 & 0.0003 & 0.0013 & 0.0034 & 0.0024 \\
		Standard deviation & 0.0409 & 0.0439 & 0.0309 & 0.0292 & 0.0245 & 0.0201 \\ \bottomrule
	\end{tabular}
	\label{tab_stat}
	\caption{The average monthly rate of return for Microsoft stock and the standard deviation for the factors}
\end{table}

Table~2 reported the correlation coefficients between the explanatory variables and excess return in the Fama-French 5 factor model. As the rule of thumb, the independent variables should not be correlated or at least the correlation between independent variables should be low. 
In three factor model (MRP, SMB
and HML) we donot observe a high correlation value between  the independent variables. 
In the three-factor model, all correlation coefficients are positive, and the lowest correlation observed between EXR and SMB.

\begin{table}[]
		\label{tab_corr}
		\centering
	\begin{tabular}{@{}lcccccc@{}}
		\toprule
		   & EXR   & MRP    & SMB    & HML       \\ \midrule
		EXR  & 1.000  & 0.569  & 0.080  & -0.341  \\
		MRP  & 0.569  & 1.000  & 0.216  & -0.190  \\
		SMB  & 0.080  & 0.216  & 1.000  & -0.250  \\
		HML  & -0.341 & -0.190 & -0.250 & 1.000    \\ \bottomrule
	\end{tabular}
	\caption{Correlation coefficients between the explanatory variables}
\end{table}
	
\begin{figure}[htb!]
		\begin{subfigure}[b]{0.5\textwidth}
			\includegraphics[width=\textwidth]{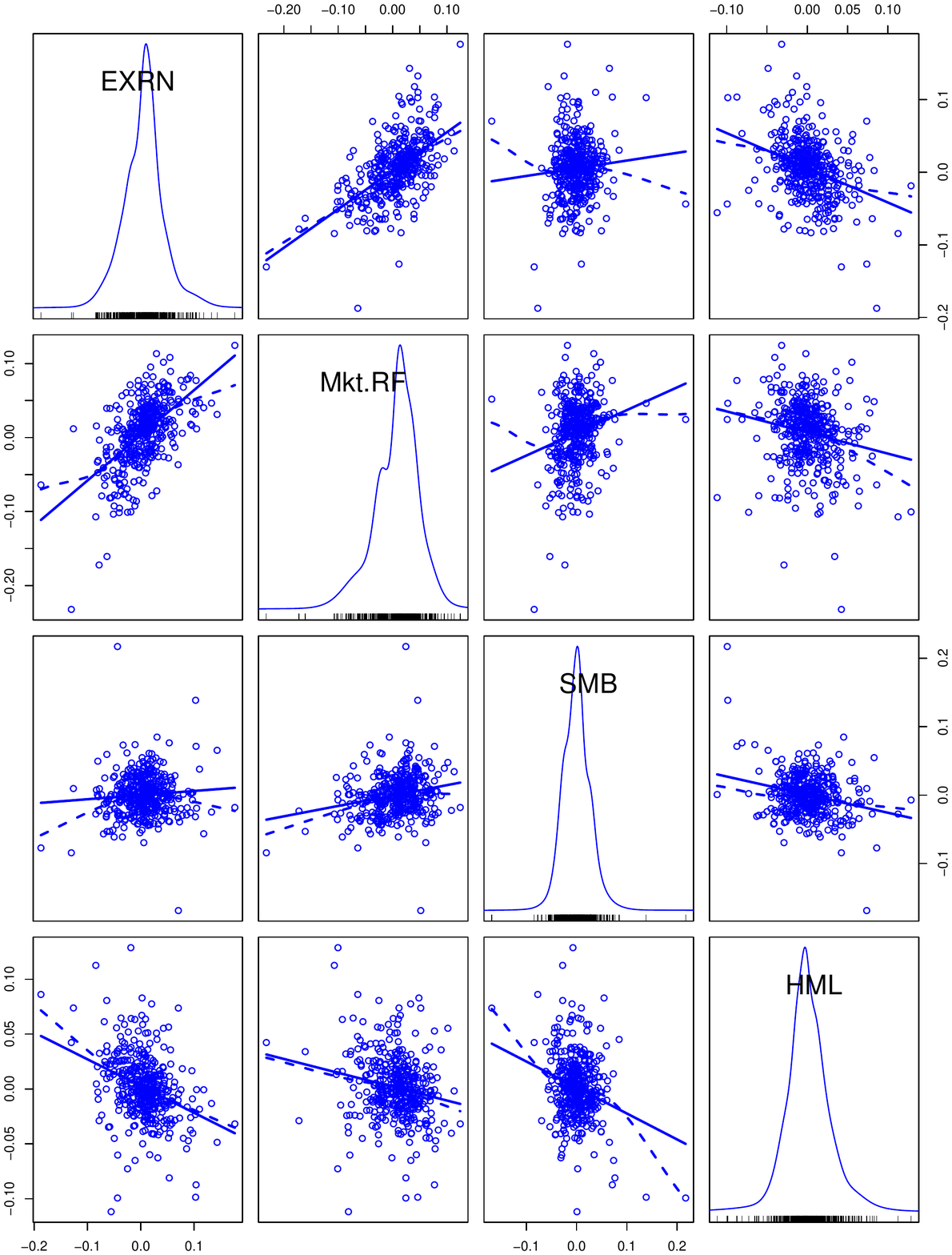}
			\label{Fig_scatter_3}
		\end{subfigure}
		\begin{subfigure}[b]{0.5\textwidth}
			\includegraphics[width=\textwidth]{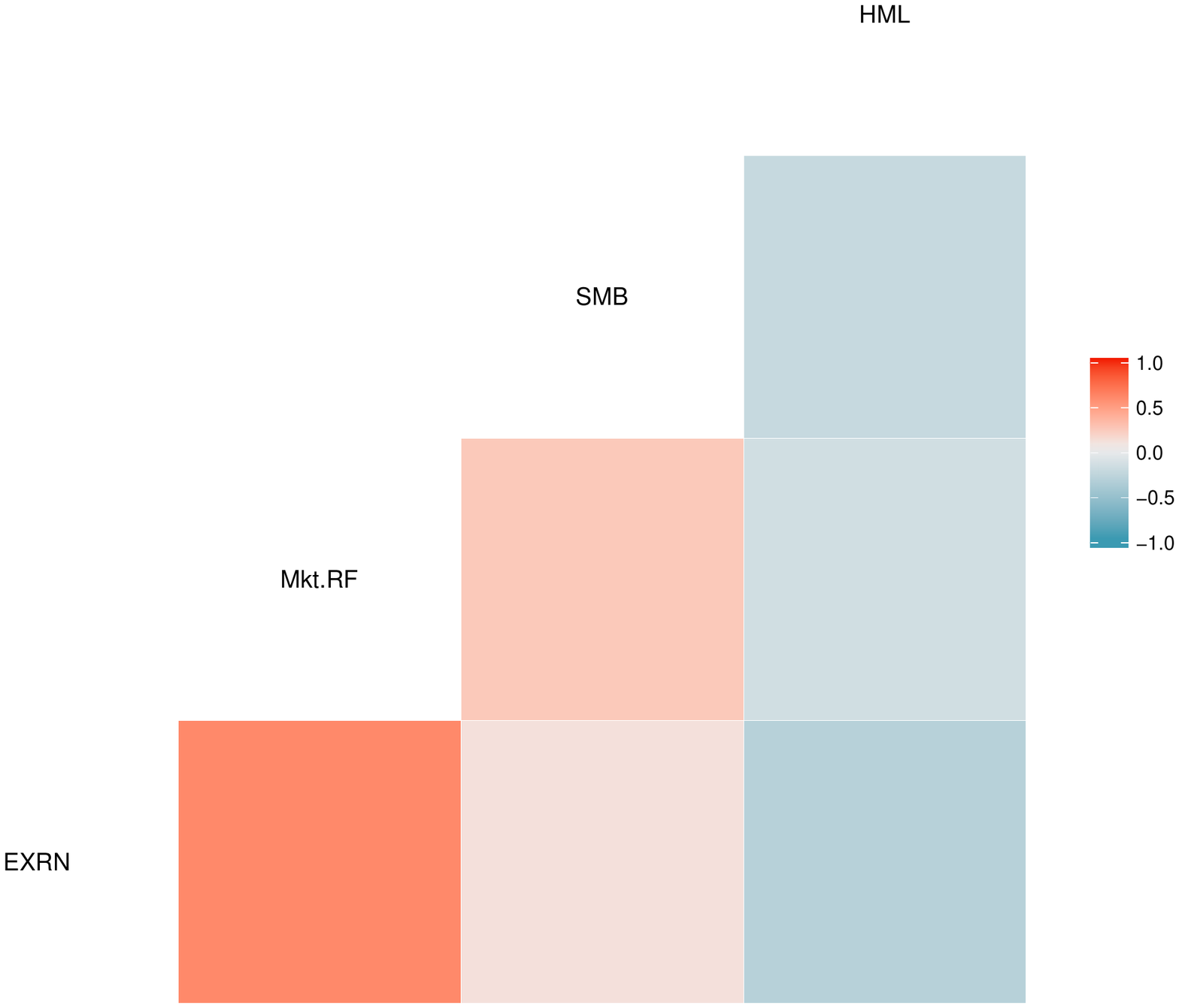}
			\label{fig_heatmap}
		\end{subfigure}
		\caption{The scatterplot matrix and the correlation heatmap for Fama-French three factor model.}
		\label{figure1}
	\end{figure}

To visualize the pairwise correlations between the variables, We plotted the scatter plot and correlation heatmap  in Figure~\ref{figure1}. From Figure~\ref{figure1} we can discover more about the nature of these relationships and the distribution's shape of the variables. Although there are outliers in the data but there is not clustering by groups in the data.   From Figure~\ref{figure1} we can quickly identify the relation between each factors in the model.

\subsection{Fama-French three-factor regression results}

	In this section we regress EXR on three factors with constant values added. The reason we add intercept to the regression model is that we need to make sure there is an intercept to confirm that the three factor model is correctly assess portfolios. If the intercept is omitted from the model, this means the model might not correctly evaluate excess return.   
	
	Table~3 shows regressions statistics for the Fama-French three factor model's explanatory variables, where three factors explain the excess returns on Microsoft stock. In the regressions model, the intercept is not statistically significant $(Pvalue =0.086)$. All factors in the Regression model to explain EXR are significant. 
	The coefficient of HML and SMB factors are negative, but the coefficient of SMB in the model is not significant at $0.01$ significant level. It is worth to note that the coefficients in regression model are close to the their correlation coefficients with the EXR. 
	The regression model suggests removing intercept and  the SMB would not hurt the explanatory power of the model if the EXR is regressed on the remaining two factors. The low value of $R^2$  $\left( 0.381\right) $ in  three factor model is a good model as  generally the  $R^2$ values in stock market are less than 50\%. Statistically significant coefficients continue to represent the mean change in the dependent variable given a one-unit shift in the independent variable. We note that in the Backward regression model, the $R^2$ of two factor model with MRP and HML as explanatory variables is $0.3806 $ not that far from the three-factor models.

	\begin{table}[]
		\centering
		\begin{tabular}{@{}lllll@{}}
			\toprule
			&Coefficients & Standard Error & t Stat & P-value        \\ \midrule
			Intercept    & 0.003          & 0.002  & 1.721   & 0.086 \\
			MRP       & 0.504          & 0.037  & 13.467  & 0.000 \\
			SMB          & -0.137         & 0.054  & -2.527  & 0.012 \\
			HML          & -0.370         & 0.057  & -6.519  & 0.000 \\ \bottomrule
		\end{tabular}
		\label{Coeff_Three}
		\caption{Regressions for the Fama-French three factor model’s explanatory variables}
	\end{table}

\begin{table}[]
		\centering
		\begin{tabular}{@{}lc@{}}
			\toprule
			Regression Statistics & \multicolumn{1}{l}{} \\ \midrule
			Multiple R            & 0.629          \\
			R Square              & 0.395          \\
			Adjusted R Square     & 0.389          \\
			Standard Error        & 0.032               \\ \bottomrule
		\end{tabular}
	\label{regressions_Three}
	\caption{Regressions Statistics for the Fama-French three factor model}
\end{table}

\begin{figure}[htb!]
	\centering
	\includegraphics[scale=0.6]{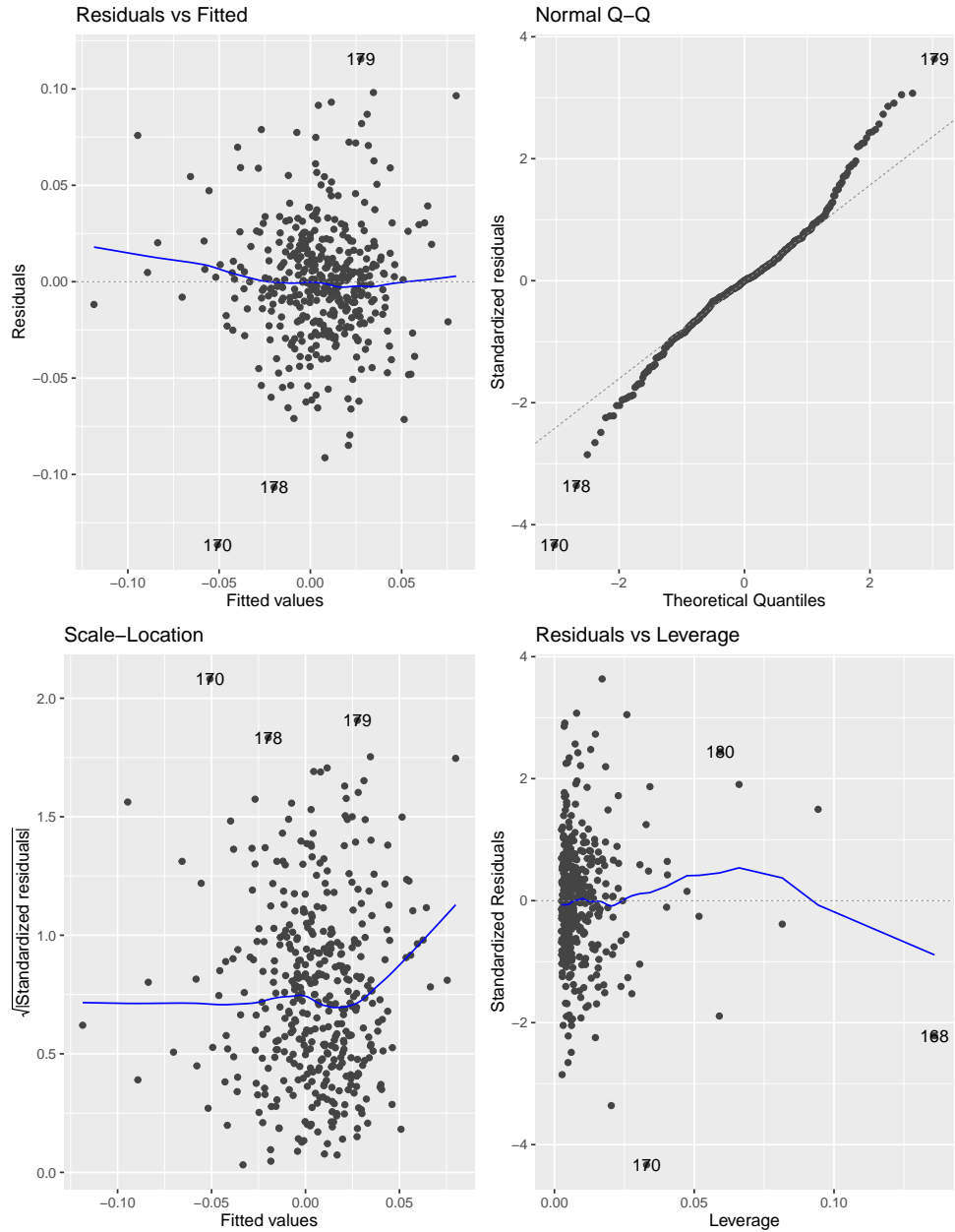}
	\caption{Diagnostic plots for three factor regression model.}
	\label{fig_diagnostic_3}
\end{figure}

Now, we go through the diagnostic plots of the model to check if our linear regression assumptions are met. The diagnostic plots are shown in Figure~\ref{fig_diagnostic_3}. The QQ plot does not show a good fit, and still, there is a problem in fitting on tails.  Q-Q plot indicates skew distributions. It seems an asymmetry with heavy tail distribution should be assumed here. The plot comes close to a straight line, except possibly for the tails, where we find a couple of residuals somewhat larger than expected. Thus, the normal assumption is not a proper assumption for the model. 

The Residuals vs. Fitted plot shows the residuals have non-linear patterns. It seems that the spread of residuals around the zero line are not equal, but the pattern is not distinct. Although the spread of residuals around the horizontal line has a mild convex trend, we can not indicate non-linear relationships and reject the assumption of equal error variances for residuals.  We can observe a few potential outliers. However, it’s probably not worthwhile to try to over-interpret this plot.

\begin{figure}[htb!]
	\centering
	\includegraphics[width=14cm, height=8cm]{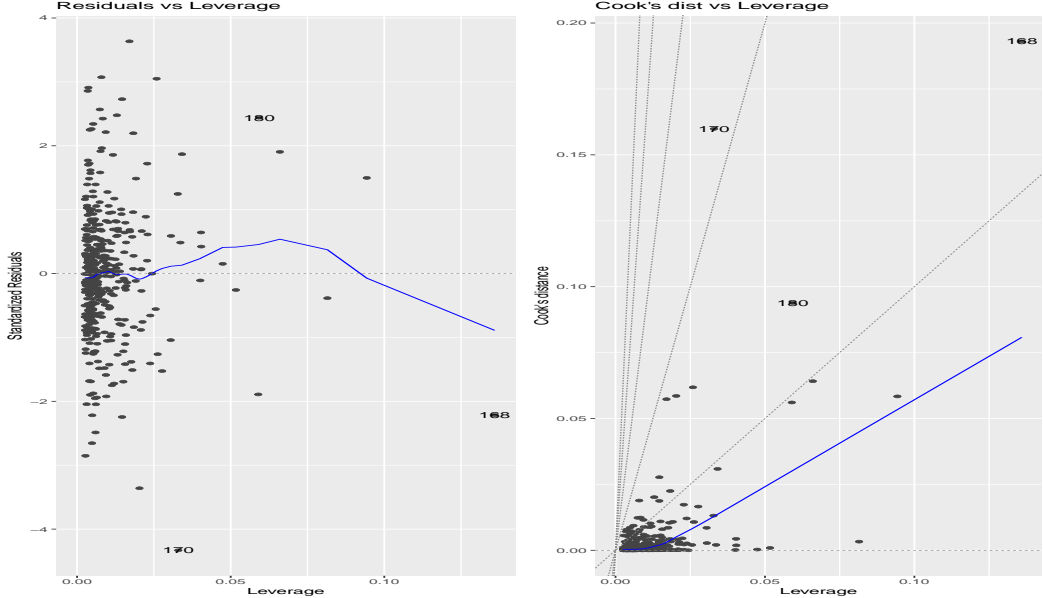}
	\caption{Diagnostic plots for three factor regression model.}
	\label{fig_Lever_3}
\end{figure}

The Scale-Location plot shows the residuals are spread equally along with the ranges of predictors. The plot shows the residuals appear randomly spread. From the Residuals vs. Leverage, we can see some influential cases. 
Although we try the new model by excluding them, the regression results did not be altered when we exclude those cases. If we exclude the outliers case from the analysis, the $R^2$ changes from $0.395$ to $0396$ with small changes in coefficients. 
To see the cases outside of a dashed line, we plot Cook's distance in Figure~\ref{fig_Lever_3}. We note that the spread of standardized residuals changes as a function of leverage in all cases. 

We can test for the assumption that the error terms  are are correlated with each other by using Durbin-Watson, d,
test statistic \citep[see][]{durbin1992}. Generally, a Durbin-Watson result between 1.5 and 2.5 indicates, that any autocorrelation in the data will not have a discernible effect on your estimates. The test statistic values for our  model is $2.0695$ indicates that we do not have an autocorrelation problem with this model. To examine normality of the residuals we run what is known as the Jarque-Bera normality test \citep[]{jarque1980}. Since we have a relatively small p-value$(\simeq 0)$ we strongly reject the null hypothesis of normally distributed errors. Our residuals are not, according to our visual examination and this test, normally distributed. 

To check  for multicollinearity, We use the Variance Inflation Factor.  A general rule of thumb is that  
$VIF>5$ is problematic. The VIF for EMR, SMB, and HML values are $1.07$, $1.11$, and $1.09$ respectively.  Thus, we are well within acceptable limits on VIF.

\subsection{Fama-French three-factor regression results for innovations}
In time-series regression analysis, the dependent behavior of explanatory variables drops the random sampling, Endogeneity, and homoscedasticity assumptions in the Gauss-Markov Theorem. 
Here to regress the EXR on three factors, we use standard econometrics models ARMA(1,1)-GARCH(1,1) with normal distribution to filter out the serial dependence and volatility clustering in each factor. Instead of studying factors and asset returns, we consider their sample innovations obtained from the time series model. The innovations are iid standardized residuals. Consequently, we have iid standardized residuals for each factor as explanatory variables in the Fama-French models. Using sample innovations instead the time series data set widely has been used in finance and economic to study the behavior of stock market and asset return process \cite[see][]{jaschke2011modelling,zhu2016extreme,shirvani2020stock}. In electrical engineering, \cite{ghaedi2016probabilistic} used this method to study hybrid electric vehicles.

Table~5 reports the correlation coefficients between the innovations of explanatory variables and excess return in the Fama-French 5 factor model. As the rule of thumb, the independent variables should not be correlated, or at least the correlation between independent variables should below. 
In the three-factor model (MRP, SMB, and HML), we do not observe a high correlation value between the independent variables. The lowest correlation coefficient is found between SMB and EXRN. 
By comparing the Tables~2 and 5, we observe the correlation coefficient between innovations are weaker than real data set. 
\begin{table}[]
	\centering
	\label{corr_inno}
	\begin{tabular}{@{}lllllll@{}}
		\toprule
		    & EXRN   & MRP    & SMB    & HML    & RMW    & CMA    \\ \midrule
		EXRN & 1.000  & 0.531  & 0.032  & -0.281 & -0.062 & -0.353 \\
		MRP  & 0.531  & 1.000  & 0.249  & -0.208 & -0.285 & -0.307 \\
		SMB  & 0.032  & 0.249  & 1.000  & -0.120 & -0.377 & -0.092 \\
		HML  & -0.281 & -0.208 & -0.120 & 1.000  & 0.035  & 0.622  \\
		RMW  & -0.062 & -0.285 & -0.377 & 0.035  & 1.000  & -0.002 \\
		CMA  & -0.353 & -0.307 & -0.092 & 0.622  & -0.002 & 1.000  \\ \bottomrule
	\end{tabular}
\caption{Correlation coefficients between the innovations of explanatory variables}
\end{table}

\begin{figure}[htb!]
	\begin{subfigure}[b]{0.5\textwidth}
		\includegraphics[width=\textwidth]{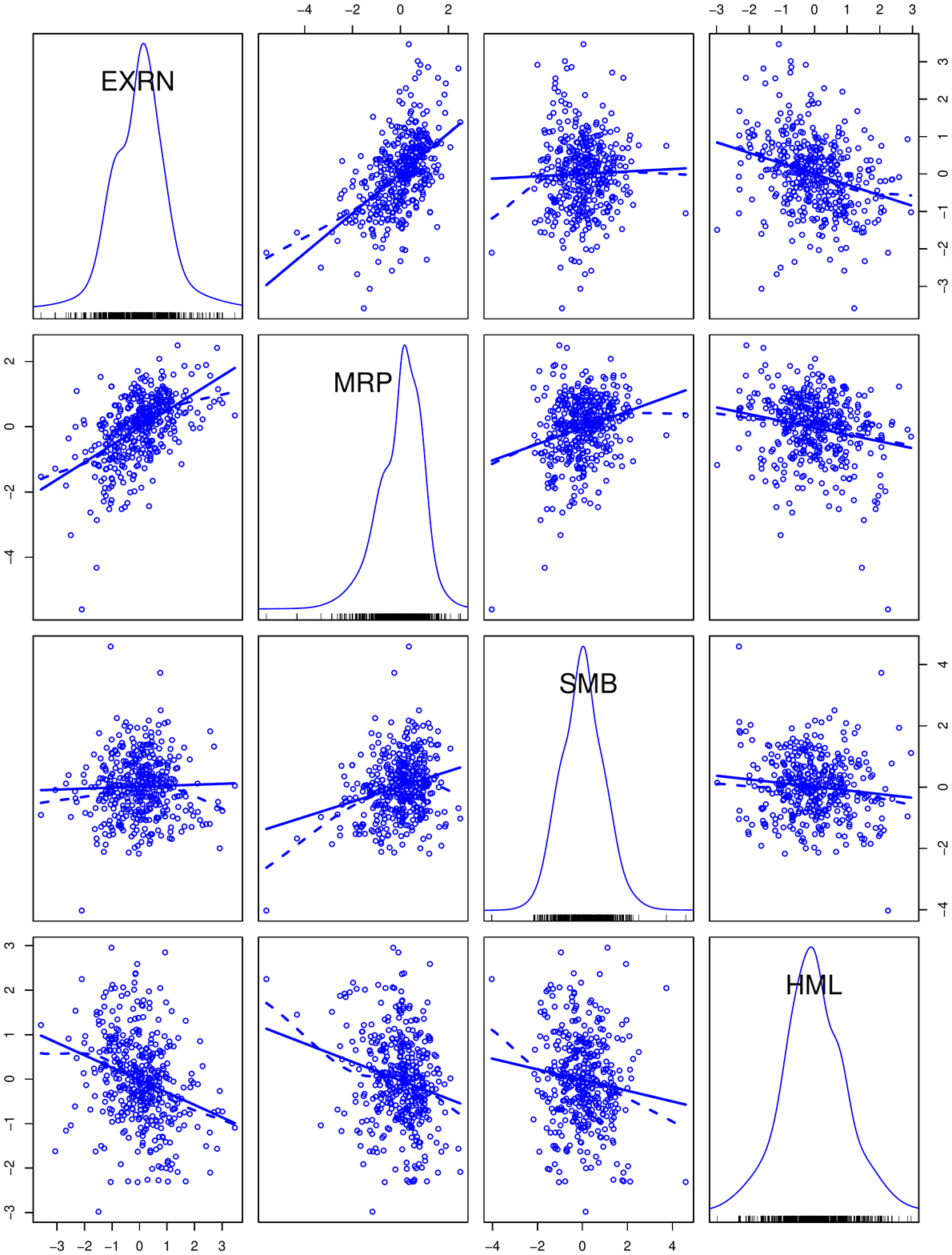}
		\label{Fig_sca_ino_3}
	\end{subfigure}
	\begin{subfigure}[b]{0.5\textwidth}
		\includegraphics[width=\textwidth]{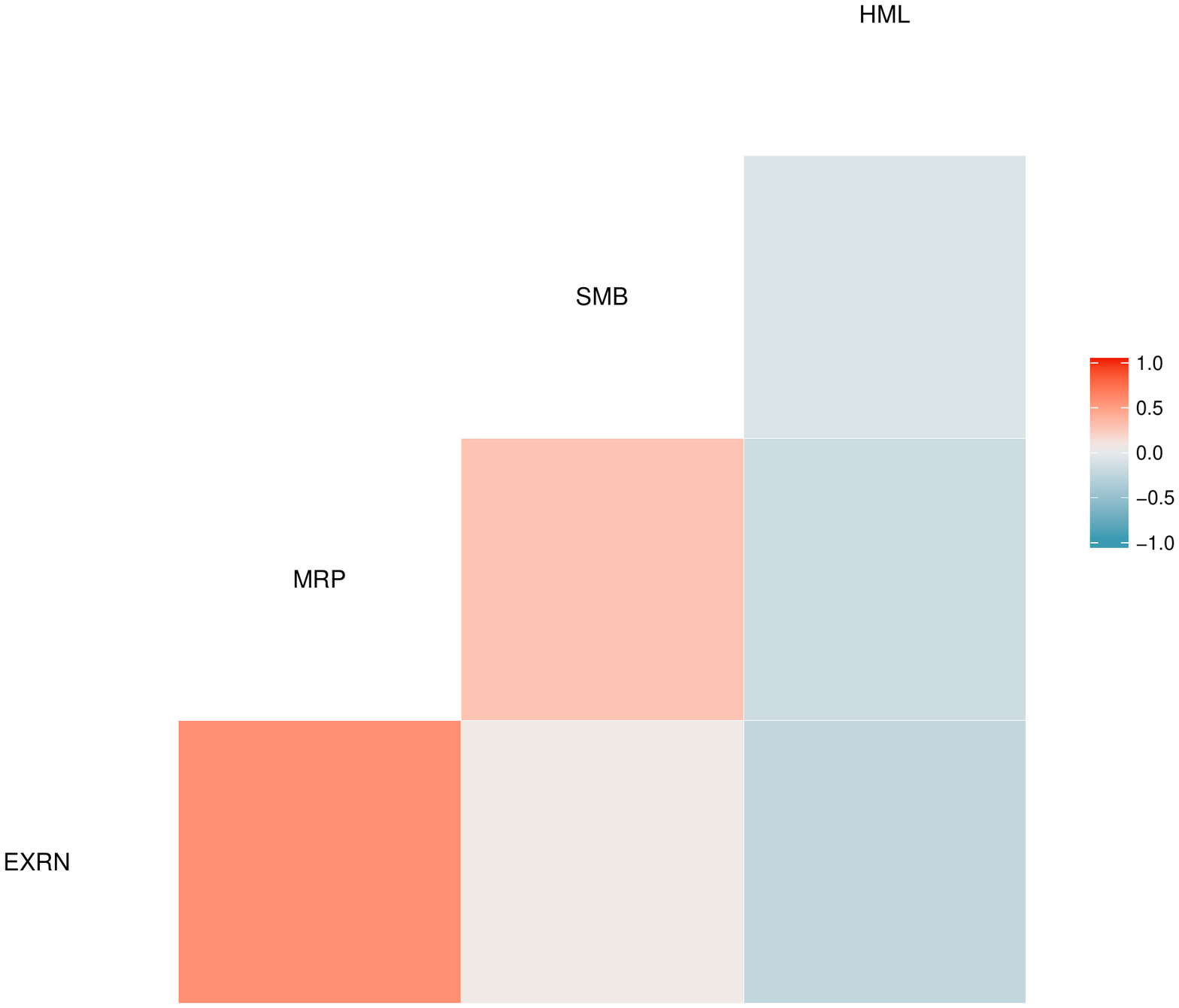}
		\label{fig_ino_heatmap}
	\end{subfigure}
	\caption{The scatterplot matrix and the correlation heatmap for innovations of Fama-French three factor model.}
	\label{figure4}
\end{figure}

The scatter plots and correlation heatmap are plotted in Figure~\ref{figure4} to visualize the pairwise correlations between the variables. From Figure~\ref{figure4}, we can discover more about the nature of these relationships and the distribution's shape of the variables. Again we observe outliers in the data, but there is not clustering by groups in the data.  From Figure~\ref{figure4}, we can quickly identify the relationship between each factor in the model.

We regress innovations of EXR on innovations of three factors with constant values added. Table~6 shows regressions statistics for the Fama-French three-factor model's explanatory variables, where the innovations of three factors explain the returns on Microsoft stock. Again the intercept is not statistically significant $(P-value =0.86)$. The difference between the innovations model with the factors model is that SMB is not signification in the Regression model in explaining EXR. The correlation coefficient of SMB with EXR confirms the insignificance of SMB in the model.  
As the regression model with factors showed an insignificant intercept and suggested removing the SMB factor, the regression model with innovations confirms it again. Hence, we can remove the SMB factor for the Fama-French model, and it does not hurt the explanatory power of the model.

\begin{table}[]
	\centering
	\begin{tabular}{@{}lllll@{}}
		\toprule
					 &   Coefficients & Standard Error & t Stat      & P-value                  \\ \midrule
		Intercept    & 0.1318    & 0.0676 & 1.950  & 0.0526 \\
		MRP          & 0.3442    & 0.0729 & 4.719  & 0.0000 \\
		SMB          & -0.1222  & 0.0659  & -1.851 & 0.0657 \\
		HML          & -0.4174   & 0.0782 & -5.338 & 0.0000 \\ \bottomrule
	\end{tabular}
\label{Coeff_Three_ino}
\caption{Regressions for the innovation of the Fama-French three factor model’s explanatory variables.}
\end{table}

$R^2$ of innovation model $(0.3521)$ is less than the factor models $(0.395)$ as we expected. Again we note that one reason for artificially high values of R-squares and adjusted R-squares for time series regressions is because the response variable and dependent variables have a trend over time. As generally, the  $R^2$ values in the stock market are less than 50\%, the variation of the EXR can be explained well by the explanatory variables.

\begin{table}[]
	\centering
	\begin{tabular}{@{}lc@{}}
		\toprule
		Regression Statistics &             \\ \midrule
		Multiple R            & 0.5933 \\
		R Square              & 0.3521 \\
		Adjusted R Square     & 0.3410 \\
		Standard Error        & 0.9051 \\ \bottomrule
	\end{tabular}
\caption{Regression statistics for the innovaitons of three factor model.}
\label{Reg_stat_ino}
\end{table}

\begin{figure}[htb!]
	\centering
	\includegraphics[scale=0.6]{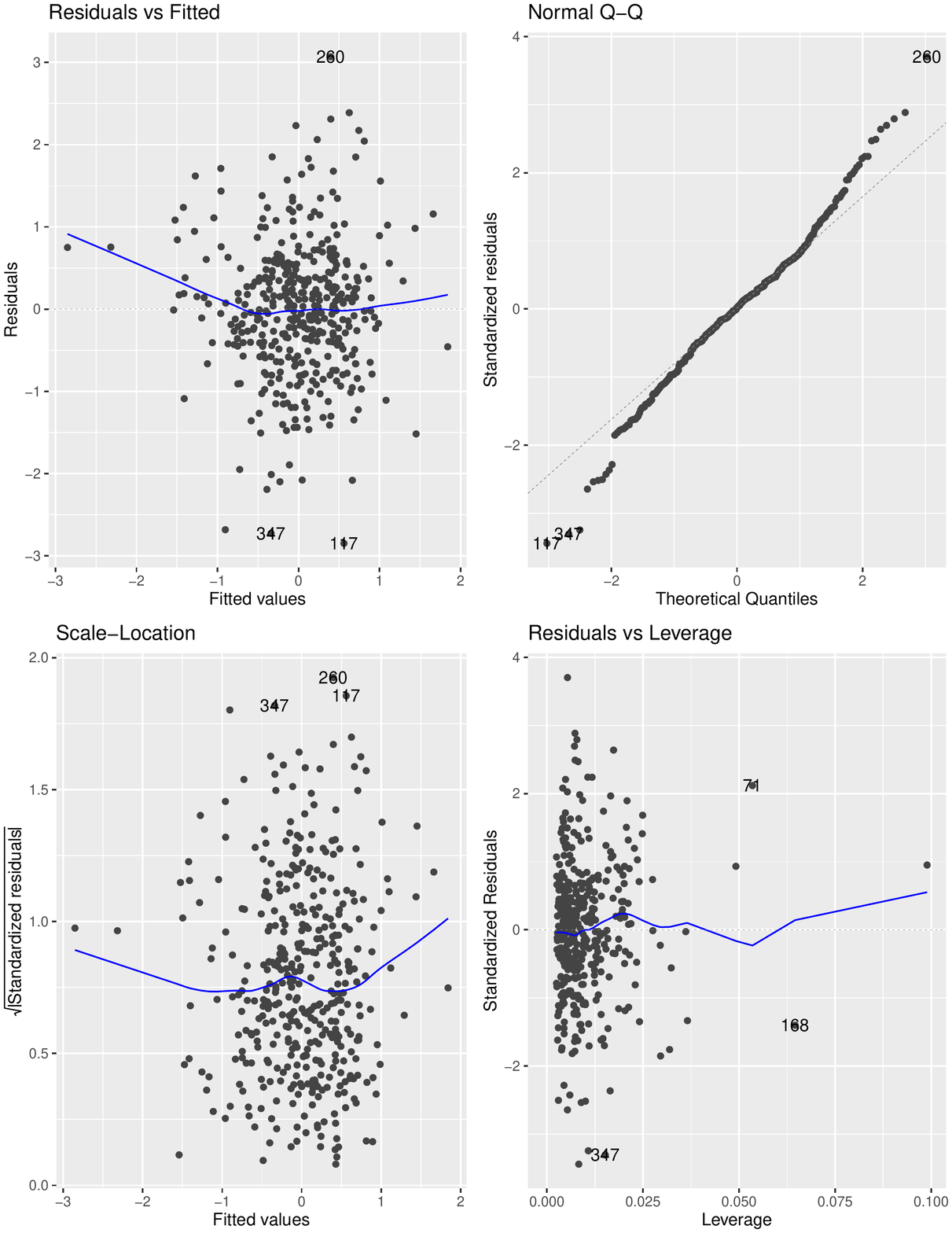}
	\caption{Diagnostic plots for three factor regression model.}
	\label{fig_diagnostic_ino3}
\end{figure}

Now, we go through the diagnostic plots for our model to check if the linear regression assumptions are met. The diagnostic plots are shown in Figure~\ref{fig_diagnostic_ino3}. The QQ plot does not show a good fit and indicates skew distributions. Still, there is a problem in fitting on tails.  It seems an asymmetry with heavy tail distribution should be assumed here. The plot comes close to a straight line, except possibly for the rear events, where we find a couple of residuals somewhat larger than expected. Thus, the normal assumption is not a proper assumption for both models. 

The Residuals vs. Fitted plot shows the residuals have non-linear patterns but not a distinct pattern. It seems that the spread of residuals around the zero line is equal. Although the spread of residuals around the horizontal line has a negative trend for the negative fitted values, again, we can not indicate non-linear relationships and rejecting the assumption of equal error variances for residuals.

Although the Scale-Location plot shows the residuals are not spread equally along with the ranges of predictors, it shows the residuals appear randomly spread. From the Residuals vs. Leverage plot, we can see some influential cases. 
We examined a new model by excluding influential cases. The regression results did not be altered when we exclude those cases. 
If we exclude the outliers from the analysis, the $R^2$ increases from $0.3521$ to $0364$ with small differences in the coefficients. To see the cases outside of a dashed line, we plot Cook's distance in Figure~\ref{fig_Lever_3}. We note that the spread of standardized residuals changes as a function of leverage in all cases.  

\begin{figure}[htb!]
	\centering
	\includegraphics[width=16cm, height=9cm]{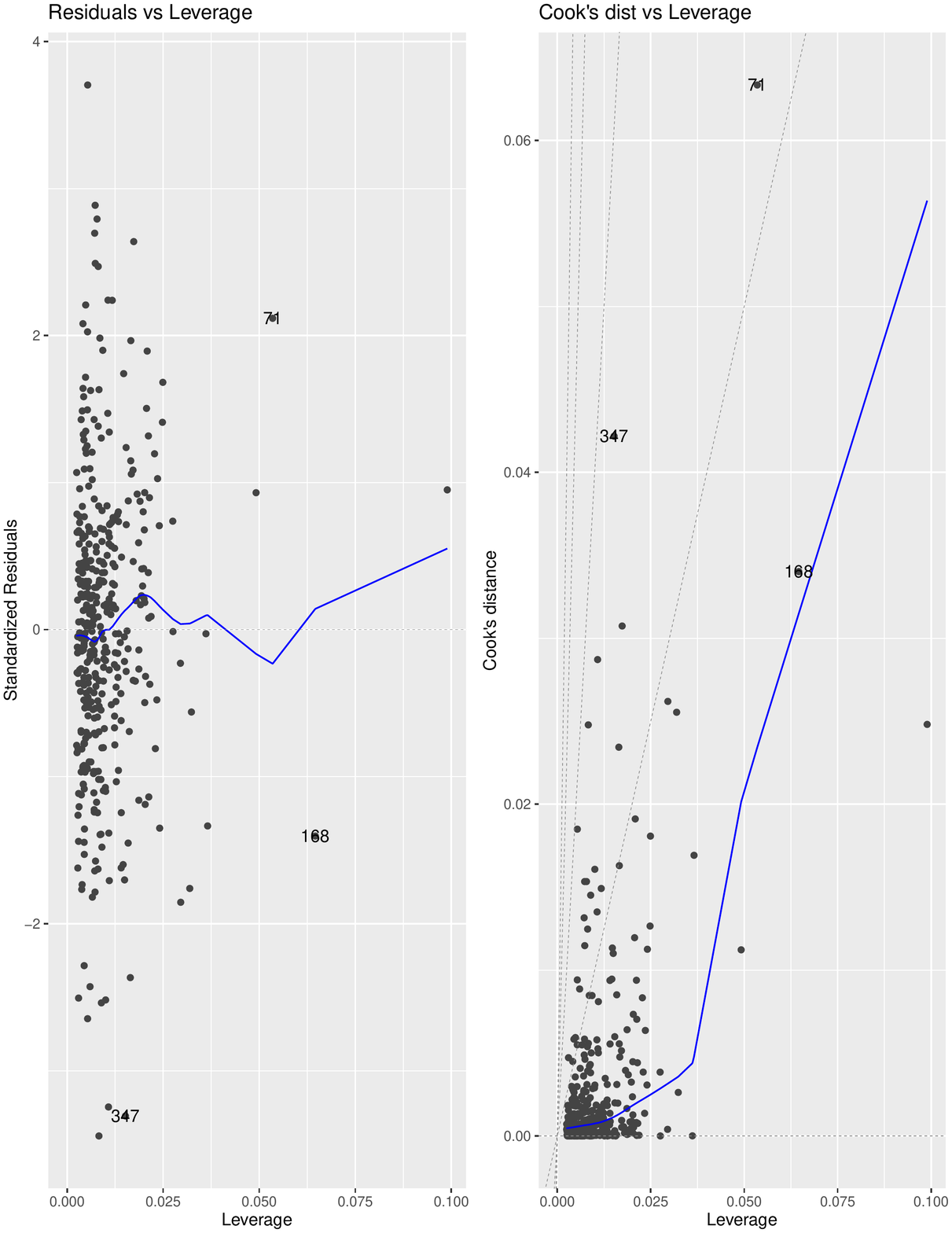}
	\caption{Diagnostic plots for three factor regression model.}
	\label{fig_Lever_ino3}
\end{figure} 

We test for the assumption that the error terms  are are correlated with each other by usin Durbin-Watson. The test statistic values for the  model is $1.9727$ indicates that we do not have an autocorrelation problem with this model. To examine normality of the residuals we perform the Jarque-Bera normality test \citep[]{jarque1980}. Since we have a relatively small p-value$(\simeq 0)$ we strongly reject the null hypothesis of normally distributed errors. Our residuals are not, according to our visual examination and this test, normally distributed. 

To check  for multicollinearity, We use the Variance Inflation Factor.  A general rule of thumb is that  
$VIF>5$ is problematic. The VIF for EMR, SMB, and HML values are $1.11$, $1.072$, and $1.051$ respectively.  Thus, we are well within acceptable limits on VIF.

	\section{Discussion and Conclusion}
	
In this study, we revisited and validated the Fama-French models in two different ways: Using the factors and asset returns in the Fama-French model and considering the sample innovations in the Fama-French model instead of studying the factors.
$R^2$  of the model with the factors is $\left( 0.381\right) $ indicates a good model in explaining the market excess return. As we expected, the $R^2 \left(=0.352 \right)$  of the model with the innovation was smaller than the model with the factors, as the time series data are correlated.  
Both regression models suggested removing the intercept, and the SMB would not hurt the explanatory power of the model. 
Using the Backward regression model, the $R^2$ of two-factor models with MRP and HML as explanatory variables are  $0.3806$ and $0.347$ for the factors and innovation models, respectively. 
The QQ plots do not show a good fit and indicate skew distributions for both models. These plots demonstrated that the normal assumption is not a proper assumption for both models. 
The Scale-Location plot shows the residuals are spread equally along with the ranges of predictors. 
The assumption that the error terms are correlated with each other tested by Durbin-Watson for both models. The test statistic values for both models demonstrated that three is no autocorrelation problem with these models. The Jarque-Bera normality test with small p-values about zero strongly rejected the null hypothesis of normally distributed errors in both models. 
We used the Variance Inflation Factor to check for multicollinearity, and results were well within acceptable limits on VIF. 

Comparing the two methods considered in this study, we suggest the Fama-French model should be consider with heavy tail distributions because

\begin{itemize}
 \item Tail behavior is relevant for regressions, including financial data,
\item  QQ plot does not validate that the choice of normal distribution as the theoretical distribution for the noise in our model,
\item In finance, the investors always try to estimate the tail, and hedge against tail risk aims to improve returns over the long-term. 	 
\end{itemize}

Finally, we note that our study was based on the Microsoft stock returns process, and we let the data speak
for itself. Future research first can be conducted for the market indices, especially for S\&P 500, Dow Jones Industrial Average index, and other stocks. It also can be generalized for other indices,  especially for Real Estate indices. Secondly, we recommend reviewing the Fama-French five-factor model by considering the sample innovations instead of studying the factors.



\end{document}